\documentclass[12pt,aps]{revtex4}   

\usepackage{amsmath}    
\usepackage{graphicx}   

\usepackage{mathrsfs}

\newcounter{con}

\begin{document}

\title{Stochastic Dynamics Toward the Steady State \\ of Self-Gravitating Systems} 

\author{Tohru Tashiro}
\email{tashiro@cosmos.phys.ocha.ac.jp}
\affiliation{Ochanomizu University}
\author{Takayuki Tatekawa}
\affiliation{Center for Computational Science and e-Systems, Japan Atomic Energy Agency}



\maketitle

\section{Introduction}

A self-gravitating system (SGS) is a system where many particles interact via the gravitational force. When we shall explain a distribution of SGS in phase space, the Boltzmann-Gibbs statistical mechanics is not useful. This is because the statistical mechanics is constructed under the condition of the additivity of the energy: as well known, the total energy of several SGSs is not equal to the sum of the energy of each system. In fact, SGS does not have a tendency to become a state characterized with a temperature.

If we use the statistical mechanics assuming that the state of the SGS with an equal mass $m$ becomes isothermal with the temperature $T$ and that the particles of the system are distributed spherically symmetrically, what kind of distribution can be obtained?
Then, the structure in phase space can be determined using the Maxwell-Boltzmann distribution. For example, the number density at a radial distance $r$ in real space is given by
\begin{equation}
n_{\rm MB}(r) \propto e^{-\frac{m}{k_{\rm B}T}\Phi(r)} \ ,
\label{nMB}
\end{equation}
where $\Phi(r)$ is the mean gravitational potential per mass generated by this whole system at $r$ and $k_{\rm B}$ is the Boltzmann constant. This potential should satisfy a relation with number density by the Poisson equation,
\begin{equation}
\frac{{\rm d}^2\Phi(r)}{{\rm d}r^2}+
\frac{2}{r}\frac{{\rm d}\Phi(r)}{{\rm d}r}
= 4\pi Gmn_{\rm MB}(r) \ ,
\label{poisson}
 \end{equation}
where $G$ is the gravitational constant. A special solution to Eq.~(\ref{nMB}) and this Poisson equation is $n_{\rm MB}(r)=k_{\rm B}T/2\pi Gm^2r^2$, known as the singular isothermal sphere~\citep{Binney87}. However, this solution has two problems: infinite density at $r=0$ and infinite total mass. Even though we solve the equations with a finite density at $r=0$, the solutions behave $\propto r^{-2}$ at a large $r$, and so we cannot avoid the infinite total mass problem. In either case, the solutions are unrealistic.

\begin{figure}[h]
\centering
 \includegraphics[scale=.75]{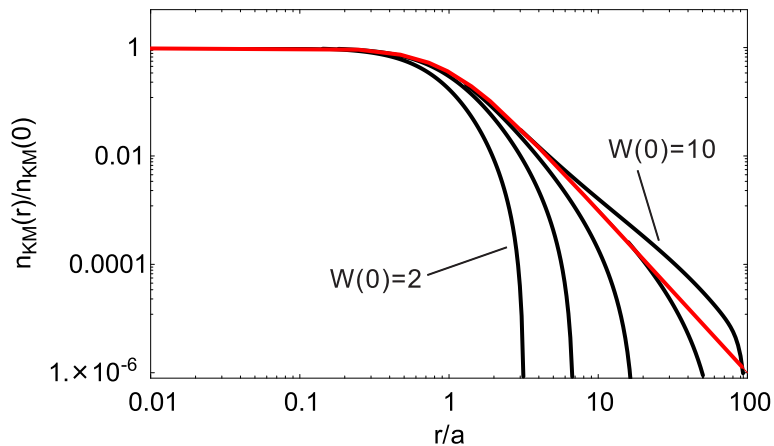}
 \caption{Black curves denote the number density of the King model $n_{\rm KM}({r})$ as a function of the radius normalized with the core radius, $r/a$, for several $W(0)$. As the curve changes from left to right, $W(0)$ gets larger from 2 to 10 in steps of 2. The red curve means the approximated formula, $1/{(1+r^2/a^2)^{3/2}}$, for larger $W(0)$.}
 \label{fig0}
\end{figure}

Of course, real examples of SGS in the universe, e.g., globular clusters and galaxies, have various structures with a finite radius. As for most globular clusters, it is known that their number densities in real space have a flat core and behave as a power law outside this core.  King interpreted these profiles by introducing the new distribution function in phase space, known as the {\em lowered Maxwellian};
\begin{equation}
f(r,v) \propto \left\{
\begin{array}{ccl}
e^{-\beta(E-m\Phi_t)} - 1 & \mbox{for} & E \le m\Phi_t \ , \\
 0                        & \mbox{for} & E > m\Phi_t \ ,
\end{array}
\right.
\end{equation}
in which $E$ is the total energy of a particle belonging to a globular cluster.
This distribution becomes zero when the total energy is greater than $m\Phi_t$, and so $\Phi_t$ can be understood as a potential energy per mass at the surface of the globular cluster. Because the velocity of the particle must be in the range of $0\le v\le \sqrt{2\left\{\Phi_t-\Phi(r)\right\}}$, the number density $n_{\rm KM}({r})$ can be obtained integrating $f(r,v)$ as
\begin{equation}
n_{\rm KM}({r}) \propto \int_{0}^{\sqrt{2\left\{\Phi_t-\Phi(r)\right\}}}{\rm d}v4\pi v^2f(r,v) \ .
\end{equation}
Moreover, using a dimensionless potential $W(r)\equiv-m\beta\left\{\Phi_t-\Phi(r)\right\}$ and integrating by parts, the number density becomes
\begin{equation}
n_{\rm KM}(r)\propto e^{W(r)}\int_{0}^{W(r)}{\rm d}\zeta e^{-\zeta}\zeta^{3/2} \ .
\label{nKM}
\end{equation}
As mentioned before, the potential energy and the number density has a relation through the Poisson equation. Thus, $W(r)$ must satisfy the following equation:
\begin{equation}
\frac{{\rm d}^2W(r)}{{\rm d}r^2}+
\frac{2}{r}\frac{{\rm d}W(r)}{{\rm d}r}
= -\frac{9}{a^2}\frac{n_{\rm KM}({r})}{n_{\rm KM}(0)} \ ,
\label{poissonW}
 \end{equation}
where $a\equiv \sqrt{9/\{4\pi Gm^2\beta n_{\rm KM}(0)\}}$ corresponds to the core radius. The number density satisfying Eqs.~(\ref{nKM}) and (\ref{poissonW}) can be calculated numerically as shown in Fig.~\ref{fig0}. This is called the King model~\citep{King1966}. When $W(0)$ is larger than about 5, the number density around the origin can be represented by the following approximation:
\begin{equation}
n_{\rm KM}({r}) \propto \frac{1}{(1+r^2/a^2)^{3/2}} \ ,
\label{nKMa}
\end{equation}
which is shown as the red curve in Fig.~\ref{fig0}.

Since King put forward this model, this number density has been applied to fitting for the surface brightness of many globular clusters, for example as in Ref.~\citep{Peterson1975,Chernoff1989,Trager1995,Lehmann1997,Meylan2001}, that is, most exponents of power law outside the core of globular clusters are similar to $-3$ which cannot be explained by the model with the isothermal assumption. But, it is not easy to see what kind of dynamics occurred in the system, because his procedure was done to the distribution function in the steady state.

So, we will construct a theory which can explain the dynamics toward such a {\em special} steady state described by the King model especially around the origin. The idea is to represent an interaction by which a particle of the system is affected from the others by a {\em special} random force described by a position-dependent intensity noise, in other words the multiplicative noise, that originates from a fluctuation only in SGS. That is, we will use a {\em special} Langevin equation, just as the {\em normal} Langevin equation with a constant-intensity noise, in other words the {additive noise}, can unveil the dynamics toward the steady state described by the Maxwell-Boltzmann distribution. However, we cannot introduce the randomness into the system without any evidence. Then, we must confirm that each orbit is random indeed. Of course, it is impossible to understand orbits of stars in globular clusters from observations. Thus, we must use numerical simulations.

From the numerical simulations of SGS, the ground that we can use the random noise becomes clear. The {\em special} Langevin equation includes additive and multiplicative noises. By using this stochastic process, we derive that non-Maxwell-Boltzmann distribution of SGS especially around the origin. The number density can be obtained through the steady state solution of the Fokker-Planck equation corresponding to the stochastic process. We exhibit that the number density becomes equal to the density profiles around the origin, Eq.~(\ref{nKMa}), by adjusting the friction coefficient and the intensity of the multiplicative noise.

Moreover, we also show that our model can be applied in the system which has a heavier particle (5-10 times as heavy as the surrounding particle).
The effect of the heavier particle in SGS, corresponding to a black hole in a globular cluster, has been studied for long time.
If the black hole is much heavier than other stars, a cusp of the density distribution appears at the center of a cluster
\citep{Peebles, Bahcall1976, Bahcall1977}.
The observations which suggest that intermediate mass black hole (IMBH, $\sim 10^2 - 10^3 M_{\odot}$) is in the globular cluster
in recent years are accomplished one after another~\citep{Clark, Newell, Djorgovski, Gebhardt, Gerssen, Noyola}.
Although these studies are very interesting, our model does not treat these situations: in our model, the heavier particle is too lighter
than IMBH.
Our model corresponds small globular cluster ($10^4$ stars) with only a stellar black hole ($\simeq 1 - 10 M_{\odot}$).

Here, note that we have reported similar results in our previous letter~\citep{Tashiro10}. In this paper, however, we demonstrate how we executed our numerical simulations. Moreover, a treatment for stochastic differential equations becomes precise, and so the analytical result derived by a different method changes a little.

This paper is organized as follows. In Sec.~\ref{GRAPE}, and Sec.~\ref{SI}, we provide brief explanations about a machine and a method we used when we did numerical simulations, respectively.
In Sec.~\ref{SND}, we investigate number densities derived from our numerical simulations where all particles of SGS with a mass $m$ and a particle with a mass $M$ interact via the gravitational force. Then, we show the densities are like that of the King model and both the exponent and the core radius are dependent on $M$.
In Sec.~\ref{FS}, forces influencing each particle of SGS are modeled. Then, using these forces, Langevin equations are constructed in Sec.~\ref{LE}. Section \ref{FPE} makes it clear that the steady state solution of the corresponding Fokker-Planck equation gives the same result with the King model. In Sec.~\ref{DC}, we discuss our results and make the relation between King's procedure and our idea clear. Section \ref{CC} gives a summary of this work.

\section{Numerical simulations of SGS using GRAPE}
\label{NS}

\subsection{GRAPE}
\label{GRAPE}

SGSs require quite long time for relaxation. Furthermore, because only attractive force is exerted on particles in SGS and the gravitational potential is asymptotically flat, we must compute interaction of all particle pairs. When we treat $N$ particles, the computation of interaction becomes $O(N^2)$ by direct approach. By these reasons, we require huge computation for numerical simulation of the evolution of SGS.

For time evolution of SGS, many improvements of algorithm and hardware have been carried out. First, we consider integrator for simulation. For long-time evolution, both the local truncation error and the global truncation error are noticed. These error occur deviation of the conservation physical quantities such as total energy. For compression of the global truncation error, symplectic integrator has been developed. The symplectic integrator conserves the total energy for long-time evolution. We apply 6th-order symplectic integrator for the time evolution of SGS. Secondly, we apply special-purpose processor for the computation of the interaction. Most of the computation of the time evolution in SGS is 2-body interaction. As special-purpose processors, GRAPE system has been developed~\citep{Sugimoto1990}. GRAPE system can compute 2-body interaction from position and mass of particles quickly. In our study, we apply GRAPE-7 chip, which consists of Field-Programmable Gate Array (FPGA) for computation of the interaction
~\citep{Kawai2006}.
GRAPE-7 chip implements GRAPE-5 compatible pipelines
\footnote{GRAPE-5 computes low-accuracy 2-body interaction. If we treat
collisional systems, i.e., the effect of 2-body relaxation cannot be neglected,
we should use high-accuracy chip such as GRAPE-6~\citep{Makino2003}.
As we will mention later,
because our simulation notices until $100~t_{ff}$, we can simulate the
systems with GRAPE-7 chip.}.
The performance of GRAPE-7 chip is approximately 100 GFLOPS and is almost equal to a processor of present supercomputers, but the energy consumption of the chip is only 3 Watts. Using sophisticated integrator and special-purpose processor, we have analyzed time evolution of SGS.

\subsection{Symplectic integrator}
\label{SI}

For time evolution, we must choose reasonable integrator for simulation.
For long-time evolution, not only the local truncation error but also
the global truncation error is noticed. For example, 4th-order Runge-Kutta
method has been applied for time evolution of physical systems~\citep{NumRec}.
Although its local truncation error is $O \left ( (\Delta t)^5 \right )$,
because its error accumulates, the global truncation error increases during
time evolution. For example, we apply 4th-order Runge-Kutta method for
the harmonic oscillation.
\begin{equation}
H = \frac{p^2}{2} + \frac{q^2}{2} \,.
\end{equation}
Using exact solutions, the orbit of the harmonic oscillation in the phase
space draws a circle. Of course, the total energy is conserved. On the
other hand, when we apply 4th-order Runge-Kutta method for time evolution,
the total energy is decreased monotonically.
\begin{equation}
H(t) = \frac{1}{2} \left[ 1- \frac{1}{72} \left ( \Delta t \right )^6
+ O \left ( \left (\Delta t \right )^7 \right ) \right]
\left ( p^2 + q^2 \right ) \,.
\end{equation}
The orbit of the harmonic oscillation in the phase space draws a spiral
and it converges to origin ($p=q=0$). When we consider long-time evolution,
4th-order Runge-Kutta method is not reasonable integrator.
If Hamiltonian of physical system is given, we can apply symplectic
integrator which based on canonical transformation
~\citep{Ruth, Feng, Suzuki, Yoshida}.
This method suppresses increase of the global truncation error.
In generic case, the symplectic integrator is implicit method.
If Hamiltonian is divided to coordinate parts and momentum parts,
the integrator becomes explicit method. The procedure of low-order integration
becomes easy more than Runge-Kutta method.
The simplest integrator is called "leap-flog method" (2nd-order integrator).
\begin{eqnarray}
p \left (t+\frac{\Delta t}{2} \right ) &=& p(t)
+ \frac{\Delta t}{2} \dot{p} \left(x(t) \right ) \,, \\
x (t+\Delta t) &=& x(t) + \Delta t \cdot p
\left (t+\frac{\Delta t}{2} \right ) \,, \\
p (t+\Delta t) &=& p \left (t+\frac{\Delta t}{2} \right )
+ \frac{\Delta t}{2} \dot{p} \left(x(t+\Delta t) \right ) \,.
\end{eqnarray}
Using leap-flog method for the harmonic oscillator,
the following equation is satisfied.
\begin{equation}
\frac{1}{2} (p^2 + q^2) + \frac{\Delta t}{2} pq = \mbox{const.} \,.
\end{equation}
Therefore the orbit in the phase space draws an oval.
The deviation from the exact
solution is suppressed. To suppress the local truncation error,
higher-order symplectic integrators have been developed. We apply
6th-order symplectic integrator for time evolution of SGS~\citep{Yoshida90}.
\begin{eqnarray}
p_i &=& q_{i-1} + c_i \Delta t \dot{p} (q_{i-1}) ~~~(1 \le i \le 8) \,, \\
q_j &=& p_{j-1} + d_j \Delta t p_j ~~~(1 \le j \le 7) \,,
\end{eqnarray}
where $p_0 = p(t), q_0 = q(t), p_8 = p(t+\Delta t), q_7=q(t+\Delta t)$.
The coefficients $c_j$, and $d_j$ are shown in Tab.~\ref{tab:Symp-6th}.
\begin{table}[h]
  \renewcommand{\arraystretch}{1.52}
\centering
\begin{tabular}{cp{0.3cm}cp{0.3cm}c}
\hline
$i$ & & $c_i$                & & $d_i$                \\
\hline
$1$ & & $0.39225680523878$   & & $0.78451361047756$   \\
$2$ & & $0.510043411918458$  & & $0.0235573213359357$ \\
$3$ & & $-0.471053385409757$ & & $-1.17767998417887$  \\
$4$ & & $0.06875316825252$   & & $1.31518632068391$   \\
$5$ & & $0.06875316825252$   & & $-1.17767998417887$  \\
$6$ & & $-0.471053385409757$ & & $0.0235573213359357$ \\
$7$ & & $0.510043411918458$  & & $0.78451361047756$   \\
$8$ & & $0.39225680523878$   & &                      \\
\hline
\end{tabular}
\caption{Coefficients of 6th-order symplectic integrator (Solution A
in~\citep{Yoshida90}).}
\label{tab:Symp-6th}
\end{table}

The symplectic integrator conserves the total energy and
the symplectic structure in generic cases. When we use $n$-th order
symplectic integrator, the local truncation error of the total energy
becomes $O\left ( (\Delta t)^{n+1} \right )$. Furthermore,
the global truncation error is not accumulated~\citep{Sanz-serna}.

In SGS, because the interaction at zero distance diverges, the local
truncation error would diverge in long-time evolution. For avoidance
of this divergence, some kind of softening parameter has been introduced
to gravitational interaction. When the nature of the pure gravity is
analyzed, the regularization procedure of interaction is required
~\citep{Kustaanheimo, Aarseth}.

\subsection{Steady number density in numerical simulation}
\label{SND}

Now, we investigate the steady number density (SND) of the SGS with a mass $m$ including a particle with a mass $M$ by numerical simulation. In particular, we show that SNDs have a core and behave as a power law outside the core.

The system is composed of $N=10000$ particles. At $t=0$, all velocities of the particles are zero and they are distributed by $n_0(r) \propto {(1+r^2/{a_p}^2)^{-5/2}} ~(0\le r \le 4{a_p})$, which is the density in real space of Plummer's solution~\citep{Binney87}. In this SGS, we put {\em another particle} with a mass $M$ in the origin at $t=0$. We shall change the mass as $M/m=1$, $5$, and $10$. Throughout this paper, we adopt a unit system where the core radius of Plummer's solution $a_p$, the initial free fall time $t_{ff}$, and the total mass $N\cdot m$ are unity.

We started the numerical simulation under these conditions.
For dynamical evolution, we used GRAPE-7 at Center for Computational Astrophysics, CfCA, of National Astronomical Observatory of Japan.
For the computation of gravitational force, we applied
Plummer's softening: the potential energy between the $i$th and $j$th particles separated by a distance $r_{ij}$ is
$-{G m^2}/{\sqrt{{r_{ij}}^2+{\varepsilon_{s}}^2}}$,
where $\varepsilon_{s}$ is the softening parameter. We set
$\varepsilon_{s} = 10^{-3}$.
For time evolution, we used  6th-order symplectic integrator~\citep{Yoshida90}.
The time step for the simulation is defined as $\Delta t = 10^{-5}$.
We carried out simulations until $t=100~t_{ff}$. During simulations,
the error in total energy was less than $0.1 \%$.

\begin{figure}[h]
\centering
 \includegraphics[scale=.78]{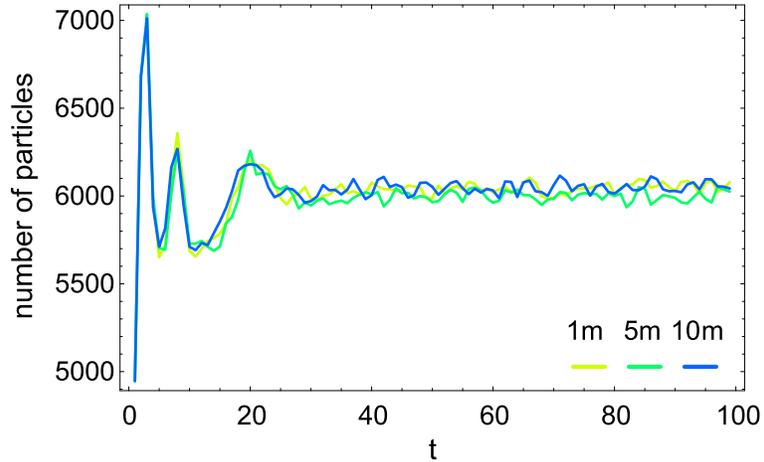}
 \caption{Change of the number of particles inside a sphere with a radius 1 in time for $M/m=1$, 5, and 10.}
 \label{fig0.5}
\end{figure}

First, most particles collapse into the origin within several $t_{ff}$. After approximately $20~t_{ff}$, the distribution becomes stable and the system reaches the steady state.
Of course, we can confirm whether the system becomes steady or not from the profile of the number density. However, furthermore we also focus on the number of particles inside a sphere.  Figure~\ref{fig0.5} shows the change of the number inside the sphere with a radius 1 in time. During the collapse, the number becomes large. After that, the number decreases, which means that many particles with positive energy evaporated from inside of the sphere, and so the number becomes about 6000 on average. For other radii, similar changes of the number in time can be seen.

\begin{figure}[tp]
  \begin{center}
    \includegraphics[scale=.75]{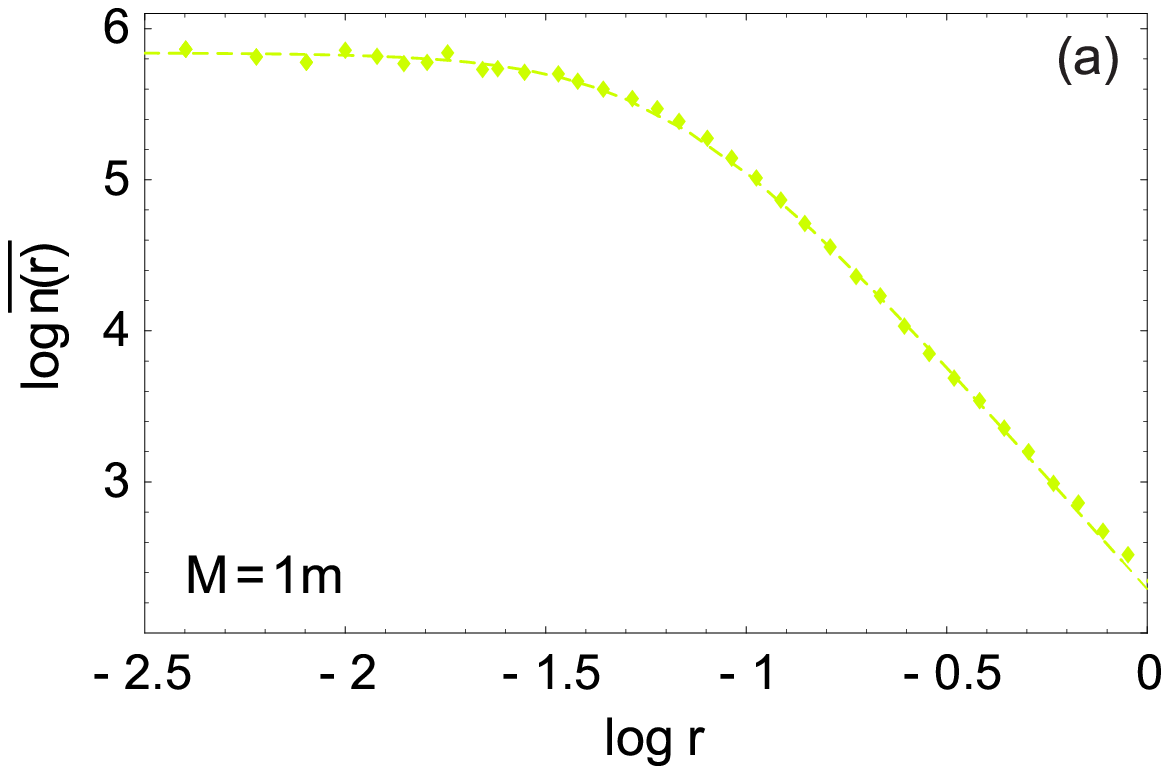}
    \mbox{}\vspace{5mm}
    \includegraphics[scale=.75]{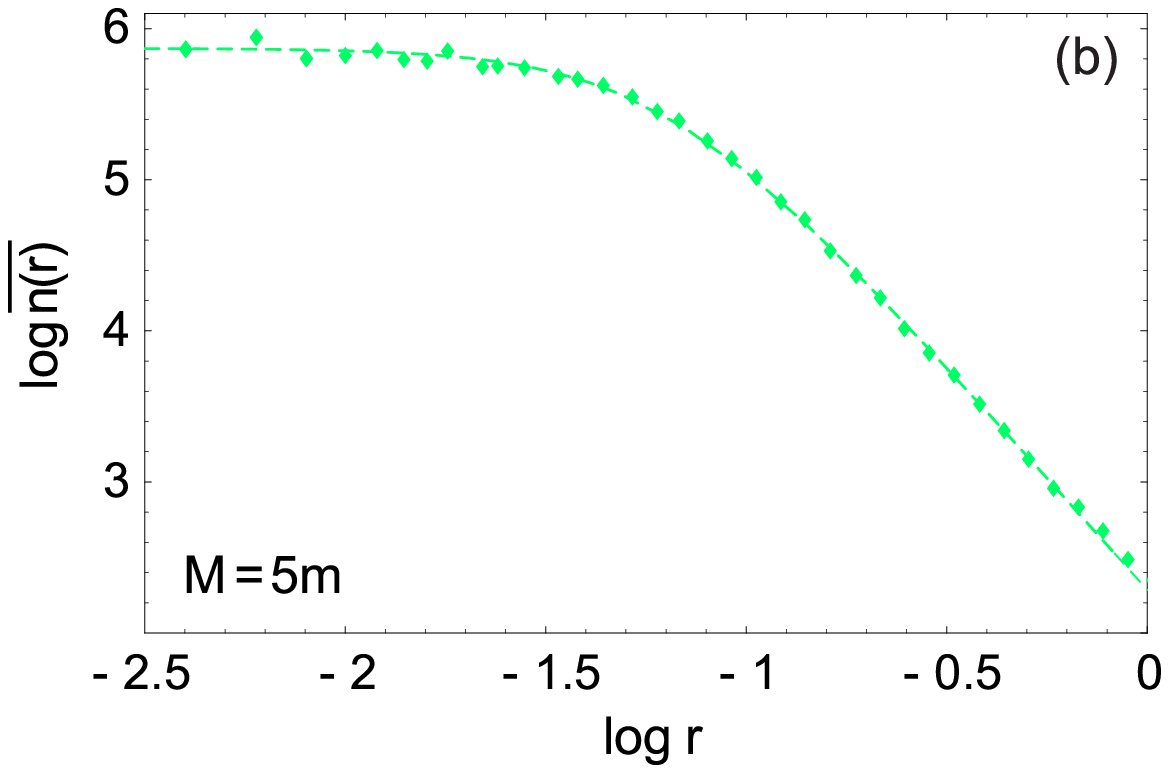}
    \mbox{}\vspace{5mm}
    \includegraphics[scale=.75]{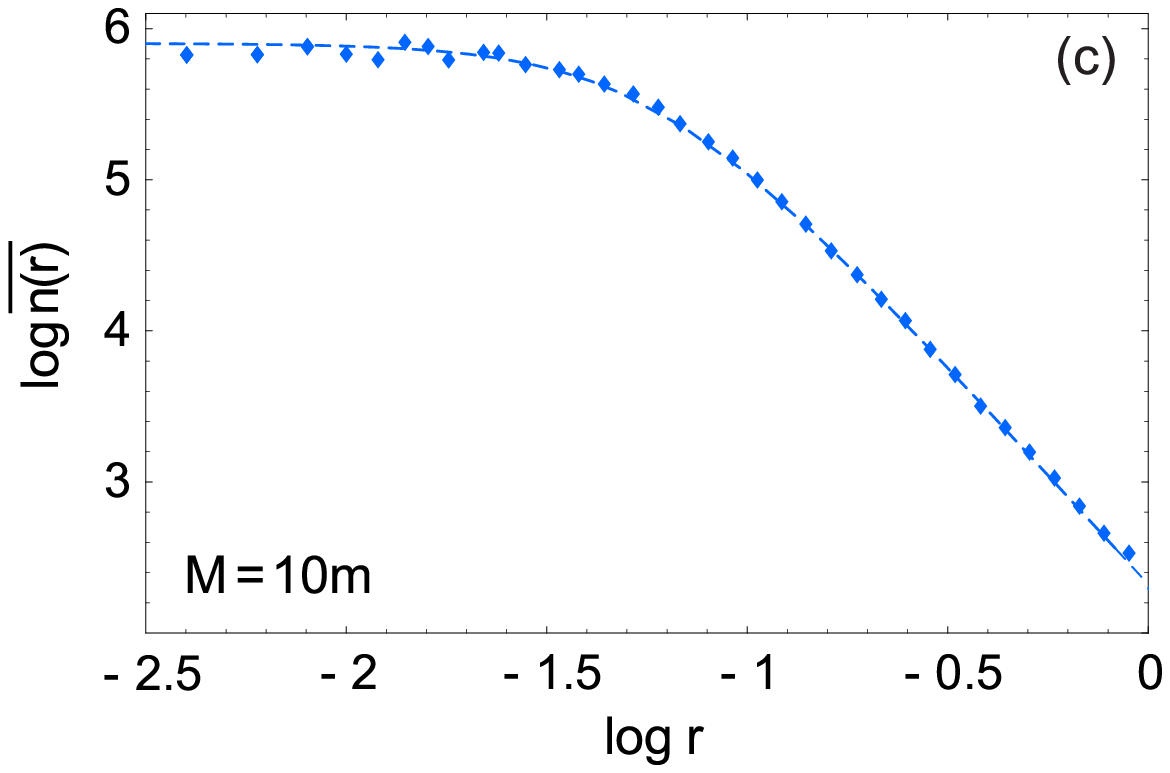}
  \end{center}
  \caption{Logarithm of the steady number density derived from our numerical simulation, $\overline{n(r)}$, as a function of $\log r$ for (a) $M/m=1$, (b) $M/m=5$, and (c) $M/m=10$. In each figure, the dashed curve and symbols denote a fitting curve $\overline{n_{\rm fit}(r)}=C/(1+r^2/a^2)^\kappa$ and the result of our numerical simulation, respectively.}
  \label{fig1}
\end{figure}

SND is calculated by taking the time average during the steady state. In Fig.~\ref{fig1}, we show the logarithm of SND as a function of $\log r$ for $M/m=1$, $5$, and $10$. For each $M$, the SND has a core and behaves as a power law at $r$ larger than the core radius.
Here, we fit SNDs around the core by
$\overline{n_{\rm fit}(r)}=C/(1+r^2/a^2)^\kappa$.
The results are summarized in Tab.~\ref{tabpara}.
For $M/m=1$ and 5, $\kappa\simeq3/2$, which is similar to the exponent of the King model.
The density at the origin $C$ increases as $M$ increases, which is simply
understood to be a result of many particles being attracted by the heavier particle.

\begin{table}[h]
  \renewcommand{\arraystretch}{1.52}
 \centering
    \begin{tabular}{cp{0.3cm}cp{0.3cm}cp{0.3cm}c}
      \hline
      $M/m$ & & $a \ \times 10^{2}$ & & $\kappa$ & & $C \ \times10^{-5}$ \\
      \hline
      1 &  & $6.20\pm0.22$  &   & $1.46\pm0.03$ &  & $7.06\pm0.14$        \\
      5 &  & $6.06\pm0.18$  &   & $1.46\pm0.02$ &  & $7.57\pm0.13$        \\
      10&  & $5.68\pm0.14$  &   & $1.43\pm0.02$ &  & $8.13\pm0.12$        \\
      \hline
    \end{tabular}
  \caption{Best-fit parameters of the function $C/(1+r^2/a^2)^\kappa$ for steady number densities shown in Fig.~\ref{fig1}}
  \label{tabpara}
\end{table}

\section{Simple model}
\label{TM}

\subsection{Forces acting on each particle of SGS}
\label{FS}

As shown in the last section, SND is the King-like profile even though the system includes the heavier particle. In this section, in order to explain these results and derive this non-Maxwell-Boltzmann distribution around the origin, we demonstrate a simple model based on stochastic process, which is quite different from the King model.

The reason why stochastic process appears in the SGS is as follows. After the collapse, the density around the origin becomes high. Thus, the particles around the region disturb the orbits of other particles repeatedly, so that their movements become random~\footnote{Generally, a particle going into a region where the gravitational potential is deep, e.g. the core of SGS, attains a high velocity. Because of many disturbances around the core, however, the mean velocity of the particle decreases, which is, naively speaking, the {\em dynamical friction}~\citep{Binney87}. Therefore, even though the heavier particle at the origin of the system makes the gravitational potential deeper, there are few particles that can escape from the core smoothly. Then more particles are drawn toward the heavier particle.}. As the time at which this disturbance occurs, we introduce the local 2-body relaxation time $t_{\rm rel}$~\citep{Spitzer1987}:
\begin{equation}
t_{\rm rel}(r) = \frac{0.065\sigma(r)^{3}}{G^2\overline{n(r)}m^2\ln(1/\varepsilon_{s})} \ ,
\end{equation}
where $\sigma(r)$ is the standard deviation of the velocity at $r$; we adopted $\ln(1/\varepsilon_{s})$ as the Coulomb logarithm.

\begin{figure}[h]
\centering
 \includegraphics[scale=.8]{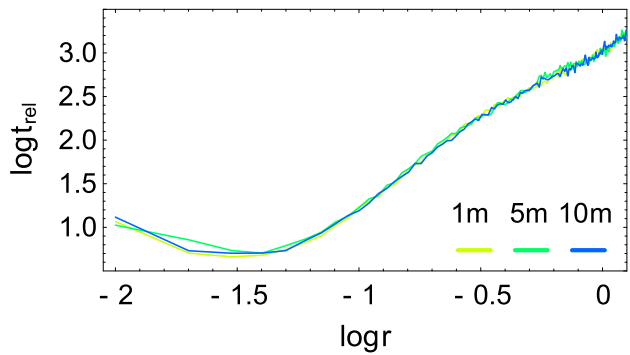}
 \caption{Logarithm of the local 2-body relaxation time $t_{\rm rel}$ as a function of $\log r$ for $M/m=1$, 5, and 10.}
 \label{fig2}
\end{figure}

Figure~\ref{fig2} shows the logarithm of $t_{\rm rel}$, which is calculated using the $\sigma(r)$ and $\overline{n(r)}$ during the steady state obtained from our numerical simulation, as a function of $\log r$. As expected, $t_{\rm rel}$ around the origin is short. Our simulation continues after the collapse at about $80~t_{ff}$, which is sufficiently longer than the $t_{\rm rel}$ around the core. As radius increases, however, $t_{\rm rel}$ becomes longer than the rest of our simulation time, which means that no stochastic motion occurs at a large $r$.
Therefore, note that our model is valid only in the neighborhood of the core.

When constructing our model, the following points are premised: the model describes the stochastic dynamics near the steady state and the mean distribution is spherically symmetric.
As is well known, the gravitational force at $r$ arising from such a spherically symmetric system depends only on the particles existing inside a sphere with a radius $r$, and this attractive force acts along the radial direction. In other words, this mean force $-F(r)$ is the gradient of the mean potential: $-F(r)=-m{\partial\Phi(r)}/{\partial r} \ (<0)$.
Indeed, $\lim_{r\rightarrow0}F(r)=0$. Hence, $F(r)$ can be expanded around the origin as
\begin{equation}
F(r)=c_0r+O(r^3) \ .
\label{f(r)}
\end{equation}
It will be clarified later that the lowest exponent must be 1 and the coefficient $c_0$ is related to the number density at the origin $C$ as
\begin{equation}
c_0 = 4\pi Gm^2C/3 \ .
\label{alphaC}
\end{equation}

For $M/m=1$, we can identify {\em another particle} together with the other particles. Contrary to this, we must consider the effect of the particle in the case $M/m\neq1$. Now, we suppose that the heavier particle exists at the origin. Then, the attractive force by this particle at $r$ is $-\mathscr{F}(r) = -GmM/r^2$. We can estimate $F(r)$ around this region as $F(r)\sim c_0r=4\pi Gm^2Cr/3$, where we used Eq.~(\ref{alphaC}). Thus, ${\mathscr{F}(r)}/{F(r)}\sim{3Mr^{-3}}/{4\pi Cm}$. This ratio becomes significant when $r\lesssim 10^{-2}$, since $C\sim10^6$ as shown in Tab.~\ref{tabpara}. Therefore, if $r$ is smaller than the radius, particles are influenced by not only $F(r)$ but also $\mathscr{F}(r)$, so that the core disappears. In fact, we have performed a numerical simulation with the heavier particle fixed at the origin, where this result is confirmed. On the other hand, Miocchi improved the King model in order to describe the steady state of a globular cluster including an IMBH and reported that the density becomes cuspy as the mass of the black hole increases~\citep{Miocchi2007}.
The nature of a globular cluster when a massive black hole
is much heavier than the surrounding star, have been studied as mentioned in Introduction.
In this case, the massive black hole stays at the center mostly. Then, a cusp of the density distribution at the center appears.
Because the heavier particles in our numerical simulation are not very heavy, the particles are not trapped at the origin. Therefore, we do not consider the effect of heavier particles explicitly and we suppose that the particles influence SGS through the density at the origin $C$: as $M$ becomes larger, it attracts more particles and $C$ increases, as shown in Tab.~\ref{tabpara}. Thus, $c_0$ is an increasing function of $M$.

It is natural to consider that the distribution fluctuates around the mean because of the many disturbances. In fact, as shown in Fig.~\ref{fig0.5}, the number of particles existing inside the sphere with a radius 1 fluctuates around the mean value.
The fluctuating part of the distribution should not be spherically symmetric, so that this produces forces along not only the radial direction, but also other directions. We assume that they are random forces and set their intensity at $r$ $2H(r)$. In addition to such random forces resulting from the fluctuating distribution, a particle at $r$ is expected to be influenced by random forces generated from neighboring particles. We set the intensity $2D$, which is independent of position.

\subsection{Langevin equations}
\label{LE}

Stochastic dynamics under the above assumptions is described by the following Langevin equations in spherical coordinates: the radial direction
\begin{equation}
ma_r + m\gamma\dot{r} = -F(r) + \sqrt{2H(r)}\eta_r(t) + \sqrt{2D}\xi_r(t) \ ,
\label{rad}
\end{equation}
the elevation direction
\begin{equation}
ma_\theta + m\gamma r\dot{\theta} = \sqrt{2H(r)}\eta_\theta(t) + \sqrt{2D}\xi_\theta(t) \ ,
\label{ele}
\end{equation}
and the azimuth direction
\begin{equation}
ma_\phi + m\gamma r\sin\theta\dot{\phi} = \sqrt{2H(r)}\eta_\phi(t) + \sqrt{2D}\xi_\phi(t)  \ ,
\label{azi}
\end{equation}
where $a_r$, $a_\theta$, and $a_\phi$ are accelerations along those directions; $\gamma$ is the coefficient of dynamical friction in the low velocity limit, independent of velocity~\citep{Binney87}. In the Chandrasekhar dynamical friction formula, the coefficient is more complicated~\citep{Binney87,Chandra1943}. However, we use the coefficient in such a limit, because the density around the core is so high that particles around there move slowly.

Now, we focus on the overdamped limit of these equations, because we have interests in the stochastic dynamics near the steady state.
In the case of the {\em normal} Langevin equation with a constant-intensity noise, we only neglect the inertial term. But, as for {\em special} Langevin equations with noises whose intensity depends on a position, the new force $-\nabla H(r)/2m\gamma$ should be considered additionally~\footnote{This force is necessary in order to interpret products in theses Langevin equations as Storatonovich ones in the corresponding stochastic differential equations. See details in Ref.~\citep{Sekimoto1999}.}. Thus, the Langevin equations in the overdamped limit becomes as follows:
\begin{eqnarray}
  \mbox{radial direction}&:& m\gamma\dot{r} = -F(r) + \sqrt{2H(r)}\eta_r(t) + \sqrt{2D}\xi_r(t) -\frac{1}{2m\gamma}H'(r) \ , \label{radOD} \\
 \mbox{elevation direction}&:& m\gamma r\dot{\theta} = \sqrt{2H(r)}\eta_\theta(t) + \sqrt{2D}\xi_\theta(t) \ , \label{eleOD} \\
 \mbox{azimuth direction} &:& m\gamma r\sin\theta\dot{\phi} = \sqrt{2H(r)}\eta_\phi(t) + \sqrt{2D}\xi_\phi(t)  \ , \label{aziOD}
\end{eqnarray}
where the prime indicates a derivative with respect to $r$.
The noises in each Langevin equation, $\xi_i(t)$ and $\eta_j(t) \ (i, j = r, \theta, \mbox{and} ~ \phi)$, are zero-mean white Gaussian and correlate only with themselves.
Indeed, the correlation function is the Dirac delta function~\footnote{It may not be natural that correlations of the random forces generated from the gravity are described by the Dirac delta function. But in this paper, for simplicity, the correlation times are assumed to be negligible. In other words, the time resolution of our simple model in the over-damped limit is assumed to be much longer than the correlation times.}.

Here, revisit the position-dependent intensity noise. We have introduced such a noise in order to represent a random force originating from the fluctuation of the distribution around the mean value which yields the mean force $-F(r)$.
Thus, the first and the second terms on the right-hand side of Eq.~(\ref{radOD}) must denote that the mean force acting along radial direction is fluctuating. As a minimal formulation describing this situation, we propose the following one:
\begin{equation}
-F(r)\{1-\sqrt{2\epsilon}\eta_r(t)\} \ ,
\label{fluMF}
\end{equation}
in which $\epsilon$ is a positive constant. This can be realized by setting
\begin{equation}
H(r)=\epsilon F(r)^2 \ .
\label{noise}
\end{equation}
Note that this {\em fluctuating mean force} is the essential feature for SGS.
Since the gravitational force is a long-range one, each particle is influenced from the whole system. The mean force is produced by the mean potential which is decided by the number density in the steady state through the Poisson equation. Obviously, this number density determines only the mean positions of the particles, and they do not remain stationary at those positions: they fluctuate. Then, the mean force also fluctuates.
$\epsilon$ indicates the extent of fluctuations.
If $\epsilon$ is 0, that is, means the mean force does not fluctuate, the stochastic dynamics of each particle is governed only by the constant-intensity random force originating from the neighboring particles. Then, the Maxwell-Boltzmann distribution is obtained as the steady solution, by which the number density of globular clusters cannot be explained as written in Introduction.

\subsection{Fokker-Planck equation and the asymptotic steady solution around the origin}
\label{FPE}

From the Langevin equations (\ref{radOD}), (\ref{eleOD}), and (\ref{aziOD}),
we obtain the Fokker-Planck equation governing the spherically symmetric probability distribution function (PDF) $P(r,t)$
{\setlength\arraycolsep{2pt}
\begin{eqnarray}
\frac{\partial}{\partial t}P(r,t) &=& \frac{D}{(m\gamma)^2}\left\{\frac{\partial^2}{\partial r^2}+\frac{2}{r}\frac{\partial}{\partial r}\right\}P(r,t) + \frac{1}{m\gamma}\frac{1}{r^2}\frac{\partial}{\partial r}r^2F(r)P(r,t) \nonumber \\
 && \!\!\!\! \mbox{}+\frac{\epsilon}{(m\gamma)^2}\left\{\frac{\partial^2}{\partial r^2}F(r)^2+\frac{2}{r}\frac{\partial}{\partial r}F(r)^2\right\}P(r,t) \ , \nonumber
\label{PDEQ}
\end{eqnarray}
}%
in which we have replaced $H(r)$ by $F(r)$ using Eq.~(\ref{noise}).
Then, the PDF with the Jacobian $\rho(r,t)\equiv4\pi r^2P(r,t)$ satisfies the following Fokker-Planck equation.
{\setlength\arraycolsep{2pt}\begin{eqnarray}
\frac{\partial}{\partial t}\rho(r,t) &=& \frac{D}{(m\gamma)^2}
\left\{\frac{\partial^2}{\partial r^2}-\frac{\partial}{\partial r}\frac{2}{r}\right\}\rho(r,t) + \frac{1}{m\gamma}\frac{\partial}{\partial r}F(r)\rho(r,t) \nonumber \\
 && \!\!\!\! \mbox{}+\frac{\epsilon}{(m\gamma)^2}\left\{\frac{\partial^2}{\partial r^2}-\frac{\partial}{\partial r}\frac{2}{r}\right\}F(r)^2\rho(r,t)
\label{rhoDEQ}
\end{eqnarray}
}%
This equation is useful when integrating with respect to $r$.

The steady state solution $\rho_{\rm st}(r)$ satisfies Eq.~(\ref{rhoDEQ}) with the left-hand side zero. By integrating the equation with respect to $r$, we have
\begin{eqnarray}
&&\left\{\frac{D}{(m\gamma)^2}+\frac{\epsilon}{(m\gamma)^2}F(r)^2\right\}\rho_{\rm st}'(r) \nonumber \\
&&\hspace{-.4cm}-\left[\frac{D}{(m\gamma)^2}\frac{2}{r}
 - \frac{\epsilon}{(m\gamma)^2}\left\{2F(r)F'(r)-\frac{2}{r}{F(r)^2}\right\}  - \frac{F(r)}{m\gamma}\right]\rho_{\rm st}(r) = \mbox{const.} \ .
\label{rhoDEQ2}
\end{eqnarray}
Now, we impose the binary condition that $P_{\rm st}(r)\equiv\rho_{\rm st}(r)/(4\pi r^2)$ and the derivative do not diverge at the origin. Then, when $r\rightarrow0$,
\begin{equation}
\rho_{\rm st}(r)=O(r^2) \ ,
\end{equation}
and
\begin{equation}
\lim_{r\rightarrow0}\rho_{\rm st}'(r)=\lim_{r\rightarrow0}4\pi(2rP_{\rm st}(r)+r^2P'_{\rm st}(r))=0 \ ,
\end{equation}
by which the constant on the right-hand side of Eq.~(\ref{rhoDEQ2}) is decided and we obtain
\begin{equation}
r\left\{{D}+{\epsilon}F(r)^2\right\}\rho_{\rm st}'(r)=
-\left[r{F(r)}\left\{2{\epsilon}F'(r)+m\gamma\right\}-{2}\left\{{D} + {\epsilon}F(r)^2\right\}\right]\rho_{\rm st}(r) \ .
\label{rho}
\end{equation}

Thus, if $F(r)$ is obtained, $\rho_{\rm st}(r)$ can also be obtained. Here, $F(r)$ relates with SND, $n_{\rm st}(r)$, through the following relation, since $-F(r)=-m\Phi'(r)$.
\begin{equation}
F'(r)+\frac{2}{r}F(r)=4\pi Gm^2n_{\rm st}(r)
\label{poissonF}
\end{equation}
Incidentally, the SND can be obtained by multiplying PDF in the steady state by total number $N$:
\begin{equation}
n_{\rm st}(r) = NP_{\rm st}(r) = \frac{N\rho_{\rm st}(r)}{4\pi r^2} \ .
\end{equation}
Therefore, equation~(\ref{poissonF}) can be represented as
\begin{equation}
F'(r)+\frac{2}{r}F(r)=\frac{GNm^2\rho_{\rm st}(r)}{r^2} \ .
\label{poissonF2}
\end{equation}

In short, $\rho_{\rm st}(r)$ and $F(r)$ are closely connected with each other through Eqs.~(\ref{rho}) and (\ref{poissonF2}).
Here, we focus on the asymptotic behaviors of them around the origin, since our model is valid around there as mentioned before. Furthermore, due to this approach, we can treat them analytically.

Assume that $F(r)$ can be expanded around the origin with the lowest exponent $k$ as follows.
\begin{equation}
F(r)=r^k\sum_{l=0}^{\infty}c_lr^l
\label{assF}
\end{equation}
Substituting this expression into Eq.~(\ref{poissonF2}), we find that $\rho_{\rm st}(r)$ can also be expanded like
\begin{equation}
\rho_{\rm st}(r)=\frac{r^{k+1}}{GNm^2}\sum_{l=0}^{\infty}c_l(k+l+2)r^l \ .
\label{assrho}
\end{equation}
After substituting both Eqs.(\ref{assF}) and (\ref{assrho}) into Eq.(\ref{rho}), we can obtain
{\setlength\arraycolsep{2pt}\begin{eqnarray}
\lefteqn{\left\{{D}+{\epsilon}r^{2k}\left(\sum_{l=0}^{\infty}c_lr^l\right)^2\right\}\frac{r^{k+1}}{GNm^2}\sum_{l=0}^{\infty}c_l(k+l+1)(k+l+2)r^l} \nonumber \\
&=&
-\left[r^{k+1}\sum_{l=0}^{\infty}c_lr^l\left\{2{\epsilon}r^{k-1}\sum_{s=0}^{\infty}c_s(k+s)r^s+m\gamma\right\}-{2}\left\{{D} + {\epsilon}r^{2k}\left(\sum_{l=0}^{\infty}c_lr^l\right)^2\right\}\right] \nonumber \\
&& \!\!\!\! \times\frac{r^{k+1}}{GNm^2}\sum_{l=0}^{\infty}c_l(k+l+2)r^l \ .
\label{rho2}
\end{eqnarray}
}

Firstly, we compare the lowest order terms on the both hand sides of Eq.~(\ref{rho2}), so that the following relation can be seen:
\begin{equation}
D\frac{r^{k+1}}{GNm^2}c_0(k+1)(k+2) = 2D\frac{r^{k+1}}{GNm^2}c_0(k+2) \ .
\end{equation}
Therefore, we can conclude that $k=1$. Secondly, compare the next lowest order terms proportional to $r^3$ and we get
\begin{equation}
D\frac{r^{2}}{GNm^2}c_1\cdot3\cdot4\cdot r = 2D\frac{r^{2}}{GNm^2}c_1\cdot4\cdot r \ ,
\end{equation}
and so $c_1=0$. Lastly, selecting only terms proportional to $r^4$ from Eq.~(\ref{rho2}), we can find
\begin{equation}
\epsilon r^2{c_0}^2\frac{r^{2}}{GNm^2}c_0\cdot2\cdot3 + D\frac{r^{2}}{GNm^2}c_2\cdot4\cdot5\cdot r^2 = -r^2c_0m\gamma\frac{r^{2}}{GNm^2}c_0\cdot3+
2D\frac{r^{2}}{GNm^2}c_2\cdot5\cdot r^2 \ ,
\end{equation}
from which the following relation can be obtained:
\begin{equation}
\frac{5}{3}\frac{c_2}{c_0} = -\frac{\epsilon {c_0}^2}{D}\left(1+\frac{m\gamma}{2\epsilon c_0}\right) \ .
\end{equation}
Without going into detail, we can see that $c_3=0$ by comparison with terms containing $r^5$. So, $\rho_{\rm st}(r)$ becomes as follows:
{\setlength\arraycolsep{2pt}\begin{eqnarray}
\rho_{\rm st}(r) &=& \frac{r^2}{GNm^2}\left(3c_0+5c_2r^2\right)+O(r^6) \nonumber \\ &=& \frac{3c_0r^2}{GNm^2}\left(1+\frac{5}{3}\frac{c_2}{c_0}r^2\right)+O(r^6) \nonumber \\
 &=& \frac{3c_0r^2}{GNm^2}\left\{1-\frac{\epsilon {c_0}^2}{D}\left(1+\frac{m\gamma}{2\epsilon c_0}\right)r^2\right\}+O(r^6) \ .
\end{eqnarray}
}
Here, if we set~\footnote{The dimension of $\sqrt{{D}/{\epsilon {c_0}^2}}$ is a length and ${m\gamma}/{2\epsilon c_0}$ is dimensionless. See ~\ref{AP}.}
\begin{equation}
a^2\equiv\frac{D}{\epsilon {c_0}^2} \ \  \mbox{and} \ \ \kappa\equiv1+\frac{m\gamma}{2\epsilon c_0} \ ,
\label{para}
\end{equation}
$\rho_{\rm st}(r)$ can be expressed around the origin like
\begin{equation}
\rho_{\rm st}(r) \sim \frac{3c_0}{GNm^2}\frac{r^2}{(1+r^2/a^2)^\kappa} \ ,
\end{equation}
which yields
\begin{equation}
n_{\rm st}(r) = \frac{N}{4\pi r^2}\rho_{\rm st}(r) \sim \frac{3c_0}{4\pi Gm^2}\frac{1}{(1+r^2/a^2)^\kappa} \ .
\label{main}
\end{equation}
Thus, we can derive the number density around the origin of SGS from the model using stochastic dynamics.

The relation~(\ref{alphaC}) is easily obtained by setting $r=0$ on Eq.~(\ref{main}), that is,
\begin{equation}
C = n_{\rm st}(0) = \frac{3c_0}{4\pi Gm^2} \ .
\end{equation}

\section{Discussion}
\label{DC}

In this section, we investigate the results derived in the preceding section and understand the roles of two noises and the heavier particle in Eq.~(\ref{main}). Additionally, we discuss the difference between the King model and our model.

As in Eq.~(\ref{para}), the exponent $\kappa$ must be larger than $1$, which does not contradict our numerical simulation shown in Tab.~\ref{tabpara}.
In order for Eq.~(\ref{main}) to correspond  completely to the King model, $\kappa=3/2$ or $\gamma = \epsilon c_0/m=4\pi Gm\epsilon C/3$ must hold.
We can regard this relation between the friction coefficient $\gamma$ and the intensity of the multiplicative noise $\epsilon$ as a kind of {\em fluctuation-dissipation relation}~\citep{Kubo}, which usually plays an important role when a stochastic process with a constant-intensity noise goes to the equilibrium state described by the Maxwell-Boltzmann distribution.

The core radius $a$ is proportional to a square root of the intensity of the additive noise $D$ owing to Eq.~(\ref{para}). Then, this intensity spreads the region where the density is almost constant. This is recognized as the effect of this noise, which makes a system homogeneous and isothermal. The existence of the core at globular clusters shows that such a diffusive effect does not disappear even for the system with long-range force. In other words, all statistical mechanical features observed in a system with short-range force, that is, {\em normal} system, does not change drastically in SGS and this effect is still universal. Our model makes it clear that the special distribution can be obtained just considering the fluctuation of mean force.

Now, let us examine the role of the mass of the heavier particle, $M$, in this system by a naive discussion. As mentioned previously, $c_0$ is an increasing function of $M$. $c_0$ exists in the denominators of $a$ and $\kappa$. Then, both values should be reduced when $M$ is increased if other parameters are independent of $M$. These theoretical expectations are consistent with our numerical results shown in Tab.~\ref{tabpara}.

How the {\em special} distribution (\ref{main}) changes if we do not consider the fluctuating mean force? The steady state solution of Eq.~(\ref{PDEQ}) with $\epsilon=0$ is
\begin{equation}
P_{\rm st}(r) \propto e^{-\frac{m^2\gamma}{D}\Phi(r)} \ .
\label{MB2}
\end{equation}
Therefore, our result goes to a singular isothermal sphere, as discussed at the beginning of this paper, by which the number density of globular clusters cannot be explained.

Here, we examine the relationship between the King model and our model. King transformed the distribution function in the phase space in order to avoid a singular isothermal sphere. In our model, we introduce the multiplicative noise into the system influenced by the mean force and the additive noise whose PDF becomes Maxwellian in the steady state, as shown in Eq.~(\ref{MB2}), so that the non-Maxwell-Boltzmann distribution (\ref{main}) is derived. In short, although these procedures seem to be different, they may have the same meaning at least around the origin.
However, we emphasize that the stochastic dynamics around there near the steady state becomes clear owing to Eqs.~(\ref{rad}), (\ref{ele}), and (\ref{azi}).

\section{Conclusion}
\setcounter{con}{\value{section}}
\label{CC}

In conclusion, the non-Maxwell-Boltzmann distribution (\ref{main}) has been obtained using the stochastic dynamics with the {fluctuating mean force} and the additive white noise. This number density can be the same as that of the King model around the origin by controlling friction coefficient and the intensity of multiplicative noise. Furthermore, our model can describe the SGS having a heavier particle.
Of course, these results are consistent with our numerical simulation.
We can say that such a stochastic dynamics occurs behind the background of the King model.
In short, the diffusive effect, which is represented by the additive noise, is universal even in SGS, and it is particular to SGS that the fluctuation of the distribution around the mean value producing the mean force makes influence on each particle of this system, which our simple model can describe.

 Finally, note that our result is available only in the neighborhood of the origin. Therefore, we must derive the density globally by further extended model and investigate the difference between the model and the King model.

\setcounter{section}{0}
\def\thesection{Appendix \Alph{section}}
\renewcommand{\theequation}{\Alph{section}.\arabic{equation}}
\setcounter{equation}{0}

\section{Dimensions of $\sqrt{{D}/{\epsilon {c_0}^2}}$ and ${m\gamma}/{2\epsilon c_0}$}
\label{AP}

From now on, $[\bullet]$ represents a dimension of $\bullet$.
Since the correlation function of the random noises $\xi_i(t)$ and $\eta_j(t) \ (i, j = r, \theta, \mbox{and} ~ \phi)$ is the Dirac delta function with argument $t$,
\begin{equation}
[\xi_i(t)] = [\eta_j(t)] = \mbox{time}^{-1/2} \ .
\end{equation}
Thus, from the expression~(\ref{fluMF}) whose dimension is a force, we can see that
\begin{equation}
[\epsilon] = \mbox{time} \ .
\end{equation}
Furthermore, from $\sqrt{2D}\xi_r(t)$ whose dimension is also a force, the dimension of $D$ can be clear like
\begin{equation}
[D] = \mbox{force}^2\cdot\mbox{time} = \mbox{mass}^2\cdot\mbox{length}^2\cdot\mbox{time}^{-3} \ .
\end{equation}
Owing to Eqs.~(\ref{f(r)}) or (\ref{alphaC}), the dimension of $c_0$ equals a force per length:
\begin{equation}
[c_0] = \mbox{force}\cdot\mbox{length}^{-1} = \mbox{mass}\cdot\mbox{time}^{-2} \ .
\end{equation}

As well known, the dimension of the damping constant, $\gamma$, is an inverse of time: $[\gamma]=\mbox{time}^{-1}$. Thereby,
\begin{equation}
\left[\sqrt{\frac{D}{\epsilon{c_0}^2}}\right] = \sqrt{\frac{\mbox{mass}^2\cdot\mbox{length}^2\cdot\mbox{time}^{-3}}{\mbox{time}\cdot\mbox{mass}^2\cdot\mbox{time}^{-4}}} = \mbox{length} \ ,
\end{equation}
and
\begin{equation}
\left[\frac{m\gamma}{\epsilon{c_0}}\right] = \frac{\mbox{mass}\cdot\mbox{time}^{-1}}{\mbox{time}\cdot\mbox{mass}\cdot\mbox{time}^{-2}} = 1 \ .
\end{equation}

\setcounter{section}{\value{con}}
\def\thesection{\arabic{section}}
\section{Acknowledgement}

We would like to thank Prof.~Masahiro Morikawa, Dr.~Osamu Iguchi, and members of Morikawa laboratory for extensive discussions.
All numerical computations were carried out on the GRAPE system at the Center for Computational Astrophysics, CfCA, of National Astronomical Observatory of Japan.
The page charge of this paper is partly supported by CfCA.
This work was supported by the Grant-in-Aid for Scientific Research
Fund of the Ministry of Education, Culture,
Sports, Science and Technology, Japan (Young Scientists (B) 21740188).



\begin{thebibliography}{99}


\bibitem[Binney \& Tremaine, 1987]{Binney87}
Binney, J. \& Tremaine, S. (1987). \emph{Galactic Dynamics},
Princeton University Press, ISBN 978-0-6910-8445-9, Princeton.

\bibitem[King, 1966]{King1966}
King, I. R. (1966). The structure of star clusters. III. Some simple dynamical models. \emph{Astron. J.}, Vol.{71}, No.1, 64-75.


\bibitem[Peterson \& King, 1975]{Peterson1975}
Peterson, C. J. \& King, I. R. (1975). The structure of star clusters. VI. Observed radii and structural parameters in globular clusters. \emph{Astron. J.}, Vol.{80}, No.6, 427-436.

\bibitem[Chernoff \& Djorgovski, 1989]{Chernoff1989}
Chernoff, D. F. \& Djorgovski, S. (1989). An analysis of the distribution of globular clusters with postcollapse cores in the galaxy. \emph{Astrophys. J.}, Vol.{339}, 904-918.

\bibitem[Trager \& King, 1995]{Trager1995}
Trager, S. C.; King, I. R. \& Djorgovski, S. (1995). Catalogue of galactic globular-cluster surface-brightness profiles. \emph{Astron. J.}, Vol.{109}, No.1, 218-241.

\bibitem[Lehmann \& Scholz, 1997]{Lehmann1997}
Lehmann, I. \& Scholz, R.-D. (1997). Tidal radii of the globular clusters M5, M12, M13, M15, M53, NGC5053 and NGC5466 from automated star counts. \emph{Astron. Astrophys.} Vol.{320}, 776-782.

\bibitem[Meylan et al., 2001]{Meylan2001}
Meylan, G.; Sarajedini, A.; Jablonka, P.; Djorgovski, S. G.; Bridges, T. \& Rich, R. M. (2001). Mayall II=G1 in M31: giant globular cluster or core of a dwarf elliptical galaxy? \emph{Astron. J.}, Vol.{122}, 830-841.



\bibitem[Peebles, 1972]{Peebles}
Peebles, P. J. E. (1972).
Star Distribution Near a Collapsed Object.
\emph{Astrophys. J.} Vol.{178}, 371-376.

\bibitem[Bahcall \& Wolf, 1976]{Bahcall1976}
Bahcall, J. N. \& and Wolf, R. A. (1976).
Star distribution around a massive black hole in a globular cluster.
\emph{Astrophys. J.} Vol.{209}, 214-232.

\bibitem[Bahcall \& Wolf, 1977]{Bahcall1977}
Bahcall, J. N. \& and Wolf, R. A. (1977).
The star distribution around a massive black hole in a globular cluster. II Unequal star masses.
\emph{Astrophys. J.} Vol.{216}, 883-907.


\bibitem[Tashiro \& Tatekawa, 2010]{Tashiro10}
Tashiro, T. \& and Tatekawa, T. (2010).
Brownian Dynamics around the Core of Self-Gravitating Systems.
\emph{J. Phys. Soc. Jpn.} Vol.{79}, 063001-1-063001-4.


\bibitem[Clark et al., 1975]{Clark}
Clark, G. W.; Markert, T. H.; Li, F. K. (1975).
Observations of variable X-ray sources in globular clusters.
\emph{Astrophys. J.} Vol.{199}, L93-L96.

\bibitem[Newell et al., 1976]{Newell}
Newell, B; Da Costa, G. S.; Norris, J. (1976).
Evidence for a Central Massive Object in the X-Ray Cluster M15.
\emph{Astrophys. J.} Vol.{208}, L55-L59.

\bibitem[Djorgovski \& King, 1984]{Djorgovski}
Djorgovski, S. \& King, I. R. (1984).
Surface photometry in cores of globular clusters.
\emph{Astrophys. J.} Vol.{277}, L49-L52.

\bibitem[Gebhardt et al., 2002]{Gebhardt}
Gebhardt, K.; Rich, R. M.; Ho, L. C. (2002).
A 20,000 $M_{solar}$ Black Hole in the Stellar Cluster G1.
\emph{Astrophys. J.} Vol.{578} L41-L45.

\bibitem[Gerssen et al., 2002]{Gerssen}
Gerssen, J. \textit{et al.} (2002).
Hubble Space Telescope Evidence for an Intermediate-Mass Black Hole in the Globular Cluster M15. II. Kinematic Analysis and Dynamical Modeling.
\emph{Astron. J.} Vol.{124}, 3270-3288.

\bibitem[Noyola et al., 2008]{Noyola}
Noyola, E.; Gebhardt, K.; Bergmann, M. (2008).
Gemini and Hubble Space Telescope Evidence for an Intermediate-Mass Black Hole in $\omega$ Centauri.
\emph{Astrophys. J.} Vol.{676}, 1008-1015.


\bibitem[Sugimoto et al., 1990]{Sugimoto1990}
Sugimoto, D.; Chikada, Y.; Makino, J.; Ito, T.; Ebisuzaki, T; Umemura, M. (1990).
 A special-purpose computer for gravitational many-body problems.
 \emph{Nature}, Vol.{345}, 33-35.

\bibitem[Kawai \& Fukushige, 2006]{Kawai2006}
Kawai, A. \& Fukushige, T. (2006). \$158/GFLOPS astrophysical N-body simulation with reconfigurable add-in card and hierarchical tree algorithm. \emph{Proceedings of the 2006 ACM/IEEE conference on Supercomputing}, No.48.

\bibitem[Makino et al., 2003]{Makino2003}
Makino, J.; Fukushige, T.; Koga, M.; Namura, K. (2003). GRAPE-6: Massively-Parallel Special-Purpose Computer for Astrophysical Particle Simulations
\emph{Pub. Astron. Soc. Japan}, Vol.{55}, 1163-1187.

\bibitem[Press et al., 2007]{NumRec} Press, W. H.; Teukolsky, S. A.; Vetterling, W. T. \& Flannery, B. P. (2007). \emph{Numerical Recipes 3rd edition}, Cambridge University Press, ISBN 978-0-5218-8068-8, Cambridge.

\bibitem[Ruth, 1983]{Ruth}
Ruth, R. (1983). A canonical integration technique. \emph{IEEE Transactions on Nuclear Science}, Vol.30, 2669-2671.

\bibitem[Feng \& Qin, 1987]{Feng}
Feng, K. \& Qin, M.-Z. (1987). The symplectic methods for the computation of Hamiltonian equations. \emph{Lecture Notes in Mathematics}, Vol.1297, 1-37

\bibitem[Suzuki, 1992]{Suzuki}
Suzuki,  M. (1992). General theory of higher-order decomposition of exponential operators and symplectic integrators. \emph{Phys. Lett. A}, Vol.{165}, 387-395.

\bibitem[Yoshida, 1990]{Yoshida90}
Yoshida, H. (1990). Construction of higher order symplectic integrators. \emph{Phys. Lett. A} Vol.{150}, 262-268.

\bibitem[Yoshida, 1993]{Yoshida}
Yoshida, H. (1993). Recent progress in the theory and application of symplectic integrators. \emph{Celes. Mech. Dyn. Astron.} Vol.56, 27-43.

\bibitem[Sanz-Serna, 1988]{Sanz-serna}
Sanz-Serna, J. M. (1988). Runge-Kutta schemes for Hamiltonian systems. \emph{BIT} Vol.{28}, 877-883.

\bibitem[Kustaanheimo \& Stiefel, 1965]{Kustaanheimo}
Kustaanheimo, P. \& Stiefel, E. (1965). Perturbation theory of Kepler motion based on. spinor regularization. \emph{J. Reine Angw. Mathematik} Vol.{218}, 204-219.

\bibitem[Aarseth, 2003]{Aarseth}
Aarseth, S. (2003). \emph{Gravitational N-body simulations}, Cambridge University Press, ISBN 978-0-5211-2153-8, Cambridge.

\bibitem[Spitzer, 1987]{Spitzer1987}
Spitzer, L. (1987). \emph{Dynamical Evolution of Globular Clusters}, Princeton University Press, ISBN 978-0-6910-8460-2, Princeton.

\bibitem[Miocchi, 2007]{Miocchi2007}
Miocchi, P. (2007). The presence of intermediate-mass black holes in globular clusters
and their connection with extreme horizontal branch stars. \emph{Not. R. Astron. Soc.} Vol.{381}, 103-116.

\bibitem[Chandrasekhar, 1943]{Chandra1943}
Chandrasekhar, S. (1943). Dynamical Friction. I. General Considerations: the Coefficient of Dynamical Friction. \emph{Astrophys. J.}, Vol.{97}, 255-262.


\bibitem[Sekimoto, 1999]{Sekimoto1999}
Sekimoto, K. (1999) Temporal coarse graining for systems of Brownian particles with non-constant temperature. \emph{J. Phys. Soc. Jpn.} Vol.{68}, 1448-1449.


\bibitem[Kubo et al., 1991]{Kubo}
Kubo, R.; Toda, M. \& Hashitsume, N (1991). \emph{Statistical Physics II: Nonequilibrium Statistical Mechanics}, Springer-Verlag, ISBN 978-3-5405-3833-2,  Berlin.







\end{thebibliography}

\end{document}